 \def\AMeV{\hbox {AMeV}}
\def\deg{$^\circ$}
\def\degg{^\circ}
\def\etal{\hbox{\it et~al.}}
\documentstyle[version2,preprint,aps]{revtex}
\begin{document}
\draft
\begin{title}
Neutrons from Multiplicity-Selected\\
Au-Au Collisions at 150, 250, 400, and 650 \AMeV\
\end{title}
\author{W. M. Zhang, R. Madey, J. Schambach\cite{JOS}, M. Elaasar\cite{ME}, }
\author{D. Keane, B. D. Anderson, A. R. Baldwin, J. W. Watson}
\begin{instit}
Department of Physics and Center for Nuclear Research,
\end{instit}
\begin{instit}
Kent State University, Kent, OH 44242
\end{instit}
\moreauthors{G. D. Westfall}
\begin{instit}
Department of Physics and Astronomy
\end{instit}
\begin{instit}
Michigan State University, E. Lansing, MI 44823
\end{instit}
\moreauthors{G. Krebs and H. Weiman}
\begin{instit}
Lawrence Berkeley Laboratory, Berkeley, CA 94720
\end{instit}
\moreauthors{J. Zhang and C. Gale}
\begin{instit}
Department of Physics
\end{instit}
\begin{instit}
McGill University, Montreal, QC, Canada, H3A--2T8
\end{instit}
\newpage
\begin{abstract}
We measured neutron triple-differential
cross sections from multiplicity-selected
Au-Au collisions at 150, 250, 400, and 650 \AMeV.
The reaction plane for each collision was estimated from the summed
transverse velocity vector of the charged
fragments emitted in the collision.
We examined the azimuthal distribution of the triple-differential
cross sections as a function of the polar angle and the neutron
rapidity. We extracted the average in--plane transverse momentum
$\langle P_x\rangle$ and the normalized
observable $\langle P_x/P_\perp\rangle$,
where $P_\perp$ is the neutron transverse momentum,
as a function of the neutron center-of-mass rapidity, and we examined
the dependence of these observables on beam energy. These
collective flow observables for neutrons, which are
consistent with those of protons plus
bound nucleons from the Plastic Ball Group, agree with
the Boltzmann--Uehling--Uhlenbeck (BUU) calculations with a
momentum--dependent interaction.
Also, we calculated the polar-angle-integrated
maximum azimuthal anisotropy ratio R from the value of
$\langle P_x/P_\perp\rangle$.

\end{abstract}
\newpage

\section{Introduction}
Predictions\cite{SM74,AA75,HS80,HS82} of
collective flow
effects in relativistic heavy-ion collisions have been
confirmed by
experiments\cite{HAG84,HAG85,PD85,KD86,KD87,PB87,HAG88,DB85,DB86,DK88,HG89,ME94,YL93}.
Components of collective flow, such as the {\it side-splash} of participants,
the {\it bounce-off} of spectators, and the out-of-plane
{\it squeeze-out}\cite{HG89,YL93} are extracted by means of
various analysis techniques.
Different methods are used to investigate
the onset of collective flow\cite{DK89,CO90,GW90,WZ90,DK91}.
Several observables have been extracted to
describe collective flow, such as the flow angle\cite{HAG84},
the average in--plane transverse
momentum\cite{PD85,KD86,KD87,DB86,DK88}, the
flow\cite{KD86}, and the scale invariant flow\cite{AB88};
however, the azimuthal correlations
of collective flow
have not been studied fully. A series of measurements were made recently with
the FOPI spectrometer at GSI and
the time projection chamber at the Lawrence Berkeley Laboratory with a
view toward examining the detailed azimuthal correlations and
the triple-differential cross sections of charged particles in the full
phase space\cite{YL93,TPCLBL}.

In this paper, we report measurements of the triple-differential
cross sections and collective flow observables
of neutrons from {\it semi-central}
Au-Au collisions at
beam energies of 150, 250, 400, and 650 \AMeV.

\section{Apparatus}
This experiment (E848H) was carried out in 1988 with the Bevalac accelerator
at the Lawrence Berkeley Laboratory.
Neutron spectra were measured
by the time-of-flight technique with (NE--102
plastic scintillator) neutron detectors
at 18 polar angles from
3\deg\ to 90\deg. All neutron detectors were 1--m long and 10--cm thick. The
widths of the neutron detectors ranged from 2.5 to 50.8 cm.
The momentum and the rapidity of the neutrons were
calculated from the measured flight time over
paths that ranged from 5.9 to 8.4 m. The flight path, the polar angle, and the
width of each neutron detector are listed in Table 1.
The properties of the neutron detectors were studied
thoroughly\cite{RC79,RM83,WZ88}; here
only the resolutions and detection efficiencies of a typical 25--cm wide
detector are listed.
Its time resolution was $390 \pm 70$ ps, and the corresponding energy
resolutions were about
4 and 38 MeV for neutrons of 150 and 650 MeV, respectively, with a
flight path of 8.4 m.
The threshold cut on the pulse height of the neutron detectors was selected
to be 16 MeVee(equivalent electron energy).
The  neutron detection efficiencies of the detector at this
threshold, as calculated with the
Monte Carlo code of Cecil {\it et al.}\cite{RC79},
varied from 7.4\% at 150 AMeV to
6.3\% at 650 AMeV. The efficiencies at this analysis threshold were insensitive
to the relatively low hardware threshold of 4 MeVee.
Particles incident on each of the 18 neutron detectors were vetoed with
a thin anticoincidence plastic scintillator in front of the neutron detector.
Simultaneous incidence of a charged particle with a neutron would
veto that neutron. The false veto rates
were estimated from auxiliary measurements
of double--hit rates of charged particles.
At the beam energy of 650 \AMeV, the rates were about 15\% for detectors close
to the beam line and a few percent for those at wide angles.
The measurement of the triple--differential cross sections was corrected for
the false veto rates.
The neutron double--hit rates for the 18 detectors were negligible
at all energies because of low neutron detection efficiencies. Because
one of the neutron detectors (at the polar angle of 7\deg) malfunctioned,
the data from this detector were discarded.

An array of 184 scintillation detectors,
4.5--m high and 5--m wide, for detection of charged particles was located
at a mean distance of
4.4 m from the target.
The size of the detectors varied from 15 cm by 15 cm to 50 cm by 60 cm.
The purpose of this array was to estimate
the azimuthal angle $\phi_R$ of the
reaction plane for each collision, and to record the multiplicity of
charged particles, which is an indication of the centrality
of the collision.  Based on a simulation with FREESCO\cite{GF82},
the double--hit rate of charged particles was estimated to be
about 20\% in an individual detector for a beam energy of
650 AMeV and smaller at lower
beam energies. Only 10\% of the charged particles
were undetected with this double--hit rate.
Neglect of this small percentage of charged particles did
not affect the measurement of the reaction plane significantly. For the
bombarding energies of 150, 250, 400, and 650 \AMeV,
the target thicknesses were 0.59, 1.1, 1.7, and 1.7 g/cm$^2$, respectively,
with (hwhm) energy spreads of 17, 14, 11, and 6\%.

\section{Determination of Reaction Plane}
We used the transverse-velocity method
reported by Fai \etal\cite{GF87}
to determine the reaction plane. This technique is an adaptation
of the transverse-momentum method of Danielewicz and Odyniec\cite{PD85}.
The detailed description of the determination is presented
elsewhere\cite{ME94,WZ90}. For Au-Au collisions reported in this paper,
we determined $\phi_R$ with charged fragments above a normalized
rapidity  $\alpha \ (\equiv Y/Y_P)_{cm} = 0.3$, where $\alpha$ is
defined as the
charged particle rapidity divided by the projectile rapidity $Y_{P}$
in the center-of-mass system;
for bombarding energies of 150, 250, 400, and 650 \AMeV,
we observed dispersions $\Delta \phi_R$ of
39.4\deg, 31.3\deg, 29.5\deg, and 25.4\deg, respectively,
with uncertainties of about 5 \%.

\section{Estimation of backgrounds}
The target-unrelated background from collisions of the Au projectiles with air
and other materials were estimated by
measurements without a target in place.
The target--induced background resulting from neutrons scattered from
the floor, ceiling, and air were measured with shadow shields
of about 1--m long between
the target and the neutron detectors. The function of the shadow shields was
to attenuate neutrons that were emitted from the target and traveled directly
to the neutron detectors.
The numbers of neutrons derived by subtracting the yields
measured with shadow shields from the yields without shadow shields
were the basic data
for extraction of the triple--differential cross sections.
The shadow shields blocked neutrons emitted from the
target, but they were also sources that generated
secondary neutrons which might hit the neutron detectors,
causing an overestimation of the background.
As an illustration of this overestimation, we display in Fig.~1
the azimuthal distribution of neutrons plus background (crosses),
background (squares), and neutrons (circles) with
rapidities $\alpha$ between
$-0.2$ to +0.2 at a polar angle of 18\deg\ for a beam energy of 650 \AMeV.
We see from Fig.~1(a) that the anisotropy in the
azimuthal distribution of
the background is stronger than that of neutrons plus background
and that the inferred azimuthal distribution of neutrons
shows negative yields (without background corrections) at azimuthal
angles $(\phi - \phi_R)$ in the vicinity of 0\deg.
The stronger anisotropy of the background and the negative yields
indicate an overestimation of the
background at azimuthal angles  $(\phi - \phi_R)$ in the vicinity of 0\deg.
We emphasize that this overestimation of the background is significant in
only a limited kinematic region
(See below).
The shadow shields located
at $\phi = 0\degg$ and 180\deg\ blocked the direct path
of neutrons from the target to the detectors,
but at the same time, provided extra material at $\phi = 0\degg$ and
180\deg\ to interact with fragments emitted
from the collisions. These secondary
interactions resulted in the production of excess neutrons.
Collisions where the difference between the azimuthal angles of
the reaction plane and the detector $\phi_R - \phi_d$ was
near 0\deg\ or 180\deg\ would produce more excess neutrons
than those with this difference away from 0\deg\ and 180\deg.
The excess neutrons produced in the collisions
with the difference $\phi_R - \phi_d$ close to
180\deg\ would not affect the background measurements significantly because
these neutrons
were produced far away from the neutron detectors
being studied; however, because excess neutrons
produced in the collisions with the difference $\phi_R - \phi_d$
close to 0\deg\ could hit the neutron detectors being studied,
these excess neutrons
could distort the azimuthal anisotropy of the background.
We made a correction
for this distortion by utilizing the observed azimuthal distributions of
both neutrons plus background and the background.
First, we used a function $\sigma_{nb}^{180}[1 + A_{nb}(\phi - \phi_R)]$
to describe the observed azimuthal distribution of neutrons plus background,
where $\sigma_{nb}^{180}$ is the cross section of neutrons plus background
at $\phi - \phi_R = \pm 180\degg$ and
$A_{nb}(\phi - \phi_R)$ is an anisotropic function which is equal to zero at
$\phi - \phi_R = \pm 180\degg$. Then,
we assumed that the correct azimuthal
distribution of the background had a form similar to that of neutrons
plus background, {\it i.e.,}
$\sigma_b^{180} [1 + A_b(\phi - \phi_R)]$, where
$\sigma_b$ is the cross section of the background at
$\phi - \phi_R = \pm 180\degg$, and $A_b$ is an
anisotropic function to be determined. The correct
background rate at $\phi - \phi_R = \pm 180\degg$ should be close to
the observed rate, as per the reasoning given above.
Finally, we assumed that the relation between
the two anisotropic functions is
$A_b(\phi - \phi_R) = K A_{nb}(\phi - \phi_R)$, where $0 \le K \le 1$ is a
constant; the cases $K = 0$ and 1
correspond a flat background and
a background with the strongest plausible correlation between the signal
and the background, respectively.
The constant $K$ was chosen to be $0.5 \pm 0.5$ when applying the correction.
This correction and the uncertainty in the correction are not significant
at low beam energies, at wide polar angles, or at high rapidities, where
$\sigma_b$ is much smaller than $\sigma_{nb}$, because this correction and
its uncertainty are proportional to $\sigma_b$; however, if $\sigma_b$ is
close to $\sigma_{nb}$, the correction and its uncertainty become significant
as in the case shown in Fig.~1(a).
The same three distributions of Fig.~1 (a) are replotted in
\hbox{Fig.~1 (b)} with
this correction for overestimating
the background; we notice from Fig.~1(b) that
the corrected neutron spectrum no longer has negative yields.

\section{Triple-Differential Cross sections}
Figure~2 shows the charged multiplicities with (solid line) and
without (dashed line) a
Au target in place for collisions at a beam energy of 650 \AMeV.
 From Fig.~2, we see that collisions with a target
are contaminated by collisions without a
target only at low charged multiplicities.
We selected collisions with a charged multiplicity $M \ge M_o =$ 30, 32, 34,
36 for beam energies of 150, 250, 400, and 650 \AMeV, respectively, to
reduce the
background from collisions of Au projectiles with air and other materials to
less than 5\% of the signal.
These selections
correspond, in a simple geometrical picture,
to maximum impact parameters of about 0.55, 0.62, 0.63, and 0.68 times the
diameter of the Au nucleus, with an uncertainty of about 10\%.
We realize that the above values of the impact parameter
do not agree with the percentages of the total
geometric cross section that we estimated earlier\cite{WZ90}.
The earlier percentages were underestimated because in that
estimate we subtracted more background than we should have.
To make a reliable estimation, we needed
a proper data sample with triggers not involving any neutrons.
Unfortunately, we did not take such a data sample in 1988; therefore,
an overestimation was hard to avoid.
We collected data necessary for the estimation of the
impact parameters later in an extension
of E848H in 1991. The impact parameters
listed above were estimated with that data sample.

The threshold cut on the neutron kinetic energy
was generally 60 MeV except for neutrons
in the seven detectors with polar angles
less than 21\deg\ for the bombarding energy of 650 \AMeV; at
this highest beam energy,
cuts on the neutron kinetic energy from 200 to 380 MeV
were needed to eliminate background contamination.

Shown in Figs.~3 to 6 are the triple-differential
cross sections
of neutrons at target-like ($-1.0 \le \alpha < -0.2$),
middle ($-0.2 \le \alpha < +0.2$), forward
($+0.2 \le \alpha < +0.7$), and
projectile-like ($+0.7 \le \alpha \le +1.2$)
rapidities, respectively. In
each of the four
figures, 36 spectra are presented in an array of four rows (for the
four bombarding energies) and nine columns (for nine out of the 17
polar angles,
selected because the neutrons detected at these angles dominate the
observed collective flow effects at the corresponding rapidity).
The closed circles and the open triangles in these figures show the results
with and without correction for the overestimation of the target--induced
background, respectively. The overestimation in the
target-like and projectile-like rapidities
is negligible as seen in Fig. 6; however, the overestimation is
serious in middle and forward rapidities, especially at high beam energies
and at small polar angles, as seen in Figs.~3 to 5.
For the beam energy of 650 \AMeV,
the spectra at 30\deg\ in Figs.~3 to 5 are void because of
electronic faults during data acquisition, and the three
plots in Fig.~4 at 9\deg, 12\deg, and 15\deg\ are empty
because of the high cuts on the energy described earlier.
The remaining 12
void spectra at the  two lowest beam energies
in Figs.~3 and 4 are a consequence of a cut of 60 MeV on neutron energy.
For each of the  rapidity bins in
Figs.~3 to 5, we notice that the
anisotropy in the azimuthal distribution at a given polar angle does
not reveal much sensitivity to the bombarding energy.
This insensitivity and its significance will be
discussed in the next section. Also, the curves shown
in Fig.~2 to 6
are the theoretical predictions from the BUU approach\cite{GB88}.
We present BUU predictions for neutrons at projectile-like rapidities
for the four beam energies of 150, 250, 400, and 650 \AMeV;
for neutrons at other rapidities, we present
predictions only for the beam enegy of 400 \AMeV\ to illustrate how
well data and theory agree with each other at rapidities away from
projectile-like rapidities.

\section{Flow observables}
We calculated the average in--plane transverse momentum
$\langle P_x\rangle$ of neutrons with normalized rapidities $\alpha$
between $-1.0$ and 1.2 based on the triple--differential cross
sections measured with much finer rapidity bins
than that shown in Figs.~3 to 6.
First, we averaged the measured in--plane transverse
momentum over the whole space and over the rapidities:
\begin{eqnarray}
	\langle P_x\rangle_{m} & = &
{\int {P(\theta,\alpha)\sin\theta \cos\phi {{d\sigma}
\over {d\theta d\cos\phi d\alpha}}
d\cos\theta d\phi d\alpha} \over
{\int {{d\sigma} \over {d\theta d\cos\phi d\alpha}}
d\cos\theta d\phi d\alpha}}\nonumber \\
& \approx &
{\sum {P(\theta=\theta_d,\alpha)\sin\theta_d \cos\phi ({{d\sigma}
\over {d\theta d\cos\phi d\alpha}})_{\theta = \theta_d}
\Delta\cos\theta \Delta\phi \Delta\alpha} \over
{\sum ({{d\sigma} \over {d\theta d\cos\phi d\alpha}})_{\theta = \theta_d}
\Delta\cos\theta \Delta\phi \Delta\alpha}},
    \label{eq:pxm}
\end{eqnarray}
where $\theta$, $\phi$, and
$\alpha$ are the polar angle, the azimuthal angle, and
the normalized rapidity of the neutrons, respectively; the neutron
momentum P is a function of $\theta$ and $\alpha$; and $\theta_d$
is the polar angle of the neutron detector.
Then, we derived the in--plane average transverse momentum
$\langle P_x\rangle$
by correcting the measured $\langle P_x\rangle_m$
for the dispersion of $\Delta \phi_R$
described in Sec.~III,
\begin{equation}
        \langle P_x\rangle  = \langle P_x\rangle_m/\cos(\Delta \phi_R).
        \label{eq:px}
\end{equation}
The uncertainty of less than 5\% in $\Delta \phi_R$ is small compared to the
other systematic uncertainties in $\langle P_x\rangle$.
Shown in Fig.~7 for the three
beam energies of 150, 250, and 400 \AMeV\ is the
average in--plane transverse momentum of neutrons versus the normalized
neutron rapidity $\alpha$. The $\langle P_x\rangle$ for
the beam energy of 650 \AMeV\ is not presented
because the relatively high and differing cuts on the energy for neutrons in
the seven detectors with small polar angles would distort the spectrum
of the average $\langle P_x\rangle$.
The uncertainties in the
triple-differential cross sections and in the dispersion
were propagated to the uncertainty in
$\langle P_x\rangle$ shown in Fig.~7.
The approximation in Eq.~(1) introduced a large uncertainty in the
calculation because of the large $\Delta\theta$ of about several
degrees; this uncertainty was folded into the systematic
uncertainties in $\langle P_x\rangle$.
The shift of the $\langle P_x\rangle$ curve
toward low rapidities, as shown in Fig.~7, was attributed to the
large systematic uncertainty in $\langle P_x\rangle$ from the large
$\Delta\theta$.
The average $\langle P_x\rangle$
decreases in the vicinity of the projectile-like rapidities.
This drop is explained by the fact that both
evaporation\cite{RM85}
and bounce-off neutrons are included in the calculation of
the average; both types of neutrons are emitted at
projectile-like rapidities and
usually have low transverse momentum.
The slope of the
average in--plane transverse momentum at
negative rapidities is steeper than that at positive rapidities because
the (60 MeV) cut on the
neutron energy rejects neutrons with low transverse momenta at negative
rapidities. The data below rapidities $\alpha$ of 0.0, $-0.1$, and
$-0.2$ are
affected by the cut on the
neutron energy for beam energies of 150, 250, and 400
\AMeV, respectively.
We extracted the slope at mid-rapidity (up to $\alpha = 0.5$)
with a linear fit to $\langle P_x\rangle$ in the region
unaffected by the cut on the neutron energy;
the uncertainty in the fit is included in the error bar.
Doss \etal\cite{KD86} defined this slope as the flow F.
Plotted in Fig.~8 with open squares are the flow F of neutrons
for the three bombarding energies;
also displayed with circles are the results
for the same three energies for protons plus
bound nucleons from the Plastic Ball Group\cite{KD86}.
The relatively
large uncertainties in the measured neutron
flow F reflect the effect on the determination of the flow F by
the shift of the $\langle P_x\rangle$ curve seen in
Fig.~7.
The results of Plastic Ball Group
are obtained without any multiplicity cuts. Our multiplicity
cuts selected the majority of the events (See Fig.~2). Neglect of
low multiplicity events would reduce the flow F by a few MeV/c, which is
small compared to the systematic uncertainties in the flow F. We did not
correct the flow F for this small reduction; instead, we included this
reduction amount in the
systematic uncertainties in the flow F.
 From Fig.~8, we see that the extracted neutron flow F
is consistent within uncertainties
with that of protons plus bound nucleons.

Also, we calculated the average of $P_x/P_\perp$, where
$P_\perp$ is the transverse momentum of neutrons.
The average $\langle P_x/P_\perp\rangle$
is plotted as a function of the neutron rapidity in Fig.~9.
The Plastic Ball Group measured the average $\langle P_x/P_\perp\rangle$
for protons plus bound nucleons at the beam energy of 200 \AMeV\cite{KD87}.
By comparing Fig.~9 in this paper and Fig.~2 in Ref.~9, we see that the
interpolated $\langle P_x/P_\perp\rangle$ of neutrons corresponding to a beam
energy of 200 \AMeV\
is almost equal to that of
protons plus bound nucleons observed in the Plastic Ball at positive
rapidities. At
projectile-like rapidities, the neutron value is a little lower than the
value for protons plus bound nucleons because of the inclusion of
evaporation of free neutrons
in our analysis.

Also, we examined another observable associated with the
azimuthal distribution about a reaction plane for the emitted neutrons,
the polar-angle-integrated maximum
azimuthal anisotropy R. The polar-angle-integrated maximum
azimuthal anisotropy R was defined and extracted by
Madey \etal\cite{RM93} for {\it semi-central} Nb-Nb
and Au-Au collisions at 400 \AMeV. The relation between the average
$\langle P_x/P_\perp\rangle$ and the
polar-angle-integrated maximum azimuthal anisotropy R can be expressed as
\begin{equation}
R = \frac {(1 + 2|\langle P_x/P_\perp\rangle|)}
           {(1 - 2|\langle P_x/P_\perp\rangle|)}.
        \label{eq:sigma}
\end{equation}
We calculated the polar-angle-integrated maximum azimuthal anisotropy R from
Eq.~(\ref{eq:sigma}); the results are shown
in Fig.~10 for the beam energies of 150, 250, and 400 \AMeV.

The difference
in the three spectra of the average
$\langle P_x/P_\perp\rangle$ and the
polar-angle-integrated
maximum azimuthal anisotropy R for the three energies is much
smaller than that in the spectra of the average in--plane transverse momentum
$\langle P_x\rangle$. In the previous section, we noticed the
insensitivity of the azimuthal
anisotropy to the beam energy; this fact is revealed clearly by
the spectra of the average
$\langle P_x/P_\perp\rangle$ in Fig.~9 and the polar-angle-integrated
maximum azimuthal anisotropy R in Fig.~10.
Previously we observed
a weak dependence of R
on the mass from Au--Au and Nb--Nb collisions\cite{METHES}.
Normalized
observables, such as the average $\langle P_x/P_\perp\rangle$ reported here,
and scaled
observables, such as the scale invariant flow\cite{AB88}, have a
weaker dependence on the beam energy and mass than observables without
scaling and normalization. This weaker dependence of scaled observables
agrees with the observations by Lambrecht \etal\cite{DL94} for neutrons from
Au--Au collisions.

\section{Theoretical Interpretation}
For theoretical interpretation, we rely here on the BUU approach\cite{GB88}
with a momentum--dependent nuclear mean field,
$U (\rho ,{\vec p} )$, as parameterized in Ref. \cite{CG90}. The
momentum-dependent interaction
is essential not only
from a theoretical standpoint\cite{JJ76},
but also  has important observable implications\cite{CG87,QP93}.
The BUU calculations simulated the experiment by
using the maximum impact
parameters that were estimated from the multiplicity selection criteria
described in Sec.~V, and imposing a neutron threshold energy of 60
MeV also as described in Sec.~V. Also, we set the incompressibility
modulus K to be 215 MeV, and we
subtracted contributions to the cross sections from composite fragments
by rejecting neutrons when the distance between the neutron and another
nucleon from the same BUU ensemble\cite{GB88} is less than a critical
distance\cite{ME94,JA85}. This critical distance was determined to be
3.3, 3.0, 3.2, and 2.7 fm
for the energies of 150, 250, 400, and 650 \AMeV, respectively,
by adjusting the BUU predictions to fit
the double--differential
cross sections $d\sigma/d\Omega$
for free neutrons. The double-differential cross
sections for the beam energies of 150, 250, 400, and 650 \AMeV\ are shown in
Fig.~11. In Fig.~11, symbols represent the data, dashed lines represent
the BUU predictions with all neutrons, and solid lines represent neutrons
that are not in clusters, as defined by the critical distance.
In general, the results obtained using coalescence prescriptions in
conjunction with one--body models like the BUU will depend on the time
during the reaction when the coalescence model is applied. A cluster
contains nucleons correlated both in coordinate space and momentum space.
It is inappropriate to apply such pictures at times that are too early
during the reaction, as two--body collisions are still too numerous and
will disrupt the cluster as well as form new ones.
Here the coalescence picture was applied at a time
corresponding to the one when the momentum distributions have  just about
relaxed to their asymptotic values; in this case, we can use a single
parameter in coordinate space because nucleons with widely differing
momenta will have separated anyway; in other words, at that time, nucleons
close together in coordinate space are in fact together in momentum space.
We verified this picture quantitatively, by spanning the momenta of the
nucleons temporarily assigned to a cluster and by rejecting those for
which it was kinematically impossible to belong to a common Fermi sphere.
This last criterion brings modifications of the straight coordinate space
picture at the level of 0.1\%. We recognize that the above simple
criterion for clustering is approximate; nevertheless, it does provide a
basis for adequate phenomenology.
We observe a weak energy dependence in
the coalescence parameter probably because we treat only the coordinate-space
part. Improvements are being made to the above approach. It is not clear
that other transport theories such as QMD provide a sound theoretical
foundation for the generation of composites.
The calculated triple--differential cross sections are shown in Fig.~6
for neutrons in the vicinity of the projectile-like rapidities for
beam energies of 150, 250, 400, and 650 \AMeV.
 From Fig.~6, we see that the BUU calculations agree generally
with the data at
400 and 650 \AMeV\ except for the most forward polar angles ({\it i.e.}, 3\deg,
5\deg, and 9\deg\ for 400 \AMeV\ and 3\deg\ and 5\deg\ for 650 \AMeV);
however, at the two lower beam energies of 150 and 250
\AMeV, the BUU calculations overestimate the triple--differential
cross sections at polar angles above 15\deg\ and underestimate below
15\deg. The comparison at the most forward angles suffers from the fact
that the experimental data at angles below 15\deg\ include  evaporation
neutrons. The failure
to obtain agreement at the two lower beam energies for polar angles
above 15\deg\ may indicate that the prescription for subtracting
composites in the BUU calculation needs to be placed on a more solid basis.

We calculated the $\langle P_x\rangle$, the flow F, and the
$\langle P_x/P_\perp\rangle$ of neutrons with the BUU theory, and the results
are presented along with the data in Figs.~7, 8, and 9,
respectively.
 From Fig.~7, we see that the
BUU calculations agree with the results
of $\langle P_x\rangle$ at middle rapidities
for the two lower energies and that
the data would agree with the BUU calculations at 400
\AMeV\ if the
spectrum of the data were shifted slightly to the center.
The BUU calculations (solid squares) of the flow F agree within uncertainties
with the data (open squares) in Fig.~8;
and in Fig.~9, the BUU calculations agree with
the data for
the average $\langle P_x/P_\perp\rangle$. The sensitivity of the flow to
the parameters of the equation-of-state has been explored in detail
by Zhang {\it et al.}\cite{JZ94}.

\section{Summary and Conclusions}
We measured neutron triple-differential
cross sections from multiplicity-selected
Au-Au collisions at 150, 250, 400, and 650 \AMeV, and extracted the neutron
collective flow observables of the average
$\langle P_x\rangle$, the flow F, and the average
$\langle P_x/P_\perp\rangle$.
The BUU calculations of the triple--differential cross sections for free
neutrons agree generally with the data at
400 and 650 \AMeV\ except for the most forward
angles where the data include evaporation neutrons.
At the two lower beam energies, the BUU calculations overestimate the
cross sections at polar angles above 15\deg. This discrepancy indicates
that the prescription for calculating free neutrons
is in fact approximate and may need
improvement for the two lower energies. It is in fact well known
that the composite ``contamination'' in the BUU grows as the bombarding
energy is lowered.  The measured neutron flow observables
agree with the calculations from the BUU theory. Also, collective flow
results for neutrons are consistent with those for protons plus bound nucleons
from the Plastic Ball Group.

\section{Acknowledgements}
This work was supported in part by the National Science Foundation under Grant
Nos.~ PHY-88-02392, PHY-91-07064,  and PHY-86-11210 and
by the U.S. Department of Energy under Grant No.~ DE-FG89ER40531 and
No.~ DEAC03-76SF00098, the Natural Sciences and Engineering Research
Council of Canada and the FCAR fund of the
Qu\'ebec Government.

\newpage
\figure{The azimuthal distribution of neutrons plus background, background,
        and neutrons emitted with midrapidities at a polar angle of
        18\deg\ from Au--Au collisions at 650 \AMeV\
        (a) with a correction and (b) without a correction for the
        overestimation of the background.\label{bkg}}
\figure{Charged particles multiplicities with (solid line) and without (dashed
       line) a target in place for 400 \AMeV\ Au beam.\label{mult}}
\figure{Triple-differential cross sections
	of neutrons with the target-like rapidities
        ($-1.0 \le \alpha < -0.2$) emitted
        at nine selected polar angles from Au--Au collisions at 150, 250,
	400 and 650 \AMeV. Closed circles and open triangles represent the
        measurements with and without a correction for the overestimation
	of background, respectively, and
        lines represent the calculations from the BUU theory
	for the beam energy of 400 \AMeV\ with an
        incompressibility modulus $K$ = 215 MeV.
        \label{TARG}}
\figure{Triple-differential cross sections
	of neutrons with the middle rapidities
        ($-0.2 \le \alpha < +0.2$) emitted
        at nine selected polar angles from Au--Au collisions at 150, 250, 400,
	and 650 \AMeV. Closed circles and open triangles represent the
	measurements with and without a correction for the overestimation
	of background, respectively, and
        lines represent the calculations from the BUU theory
	for the beam energy of 400 \AMeV\ with an
        incompressibility modulus $K$ = 215 MeV.
        \label{NMID}}
\figure{Triple-differential cross sections
	of neutrons with the forward rapidities
        ($+0.2 \le \alpha < +0.7$) emitted
        at nine selected polar angles from Au--Au collisions at 150, 250, 400,
	and 650 \AMeV. Closed circles and open triangles represent the
	measurements with and without a correction for the overestimation
	of background, respectively, and
        lines represent the calculations from the BUU theory
	for the beam energy of 400 \AMeV\ with an
        incompressibility modulus $K$ = 215 MeV.
        \label{MIDD}}
\figure{Triple-differential cross sections
	of neutrons the projectile-like rapidities
	($+0.7 \le \alpha \le +1.2$)
        at nine selected polar angles from Au--Au collisions at 150, 250, 400,
	and 650 \AMeV. Closed circles and open triangles represent the
	measurements with and without a correction for the overestimation
	of background, respectively, and
        lines represent the calculations from the BUU theory with an
        incompressibility modulus $K$ = 215 MeV.
        \label{PROJ}}
\figure{Average in--plane transverse momentum of neutrons
        as a function of the rapidity for
         150, 250, and 400 \AMeV\ Au-Au collisions. Symbols represent the
         data and lines represent the calculations from the BUU theory
         with an incompressibility modulus $K$ = 215 MeV
	 \label{PX}}
\figure{Flow of neutrons (open squares) and protons plus
bound nucleons (circles)
from Au--Au collisions as a function of
the beam energy. The solid squares represent the calculations from the BUU
theory.\label{flow}}
\figure{Normalized average $\langle P_x/P_\perp\rangle$ of neutrons
       as a function of the rapidity for
         150, 250, and 400 \AMeV\ Au-Au collisions. Symbols represent the
         data and lines represent the calculations from the BUU theory.
	 \label{COS}}
\figure{Polar-angle-integrated maximum azimuthal anisotropy R of neutrons
       as a function of the rapidity for
         150, 250, and 400 \AMeV\ Au-Au collisions.
	 \label{ANISO}}
\figure{Double-differential cross sections
	for neutrons emitted from Au--Au collisions at
        150, 250, 400, and 650\AMeV.
        Symbols represent the data and lines represent
        the calculations from the BUU theory: dashed lines represent
        all neutrons and solid lines represent neutrons that
        are not in clusters.
        \label{DDXSEC}}

\newpage
\begin{table}
\caption{Flight path, polar angle, and width of each neutron
detector}\label{neudet}
\begin{tabular}{cccc}
Detector     & Polar      & Flight & Width  \\
number       & angle      & path   &        \\
             & (deg)      & (m)    & (cm)   \\
\tableline
  1          &   $ 3$    & $8.32$ &  2.5     \\
  2          &   $ 5$    & $8.31$ &  2.5     \\
  3          &   $ 7$    & $8.38$ &  12.5     \\
  4          &   $ 9$    & $8.38$ &  12.5     \\
  5          &   $12$    & $8.36$ &  25.4    \\
  6          &   $15$    & $8.36$ &  25.4    \\
  7          &   $18$    & $8.37$ &  25.4    \\
  8          &   $21$    & $8.36$ &  25.4    \\
  9          &   $24$    & $8.38$ &  25.4    \\
 10          &   $27$    & $8.38$ &  25.4    \\
 11          &   $30$    & $8.35$ &  25.4    \\
 12          &   $36$    & $8.34$ &  50.8    \\
 13          &   $45$    & $8.33$ &  50.8    \\
 14          &   $54$    & $7.92$ &  50.8    \\
 15          &   $63$    & $7.42$ &  50.8    \\
 16          &   $72$    & $6.91$ &  50.8    \\
 17          &   $81$    & $6.41$ &  50.8    \\
 18          &   $90$    & $5.91$ &  50.8    \\
 \end{tabular}
 \end{table}
\end{document}